\documentclass[twocolumn,english]{IEEEtran}
\usepackage[T1]{fontenc}
\usepackage[latin9]{inputenc}
\usepackage{float}
\usepackage{subscript}

\makeatletter

\providecommand{\tabularnewline}{\\}
\floatstyle{ruled}
\newfloat{algorithm}{tbp}{loa}
\providecommand{\algorithmname}{Algorithm}
\floatname{algorithm}{\protect\algorithmname}

\newenvironment{lyxcode}
	{\par\begin{list}{}{
		\setlength{\rightmargin}{\leftmargin}
		\setlength{\listparindent}{0pt}
		\raggedright
		\setlength{\itemsep}{0pt}
		\setlength{\parsep}{0pt}
		\normalfont\ttfamily}%
	 \item[]}
	{\end{list}}

\usepackage{url}
\usepackage{hyperref}

\makeatother

\usepackage{babel}
\begin{document}
\title{Estimating the Increase in Emissions caused by AI-augmented Search}
\author{\IEEEauthorblockN{Wim Vanderbauwhede} \IEEEauthorblockA{School of Computing Science\\ University of Glasgow\\ Glasgow, UK\\ Email: wim.vanderbauwhede@glasgow.ac.uk}}
\maketitle
\begin{abstract}
AI-generated answers to conventional search queries dramatically increase
the energy consumption. By our estimates, energy demand increase by
60-70$\times$. This is a based on an updated estimate of energy consumption
for conventional search and recent work on the energy demand of queries
to the BLOOM model, a 176B parameter model, and OpenAI's GPT-3,
which is of similar complexity.
\end{abstract}

\section{Introduction}

The new trend in search engines, to provide an AI-generated answer
to the search query, has a considerable impact on the energy consumption
and therefore CO2 emissions per query. To illustrate the impact of
AI augmented search queries more clearly, I compare the energy consumption
and emission of a query to Google's BLOOM model with that of a conventional
Google search-style query. If all search queries are replaced by AI-augmented
queries, what does that mean for energy consumption and emissions?

\section{Google search energy and emissions}

In 2009, The Guardian published an article about the carbon cost of
Google search \cite{Hickman2009}. Google had posted a rebuttal \cite{Google2009}
to the claim that every search emits 7 g of CO\textsubscript{2} on
their blog. What they claimed was that, in 2009, the energy cost was
0.0003 kWh per search, or 1 kJ. That corresponded to 0.2 g CO\textsubscript{2},
and I think that was indeed a closer estimate.

This number is still often cited but it is entirely outdated. In the
meanwhile, computing efficiency has rapidly increased \cite{Masanet202}:
Power Usage Effectiveness (PUE, metric for overhead of the data centre
infrastructure) dropped by 25\% from 2010 to 2018; server energy intensity
dropped by a factor of four; the average number of servers per workload
dropped by a factor of five, and average storage drive energy use
per TB dropped by almost a factor of ten. Google has released some
figures about their data centre efficiency REF that are in line with
these broad trends. It is interesting to see that PUE has not improved
much in the last decade.

Therefore, with the current AI hype, I wanted to revise that figure
from 2009. Three things have changed: the carbon intensity of electricity
generation has dropped \cite{OWID2023}, server energy efficiency
has increased a lot, and PUE of data centres has improved \cite{Google2023}.
Combining all that, my new estimate for energy consumption and the
carbon footprint of a Google search is 0.00004 kWh and 0.02 g CO\textsubscript{2}
(using carbon intensity for the US). According to Masanet\cite{Masanet202},
hardware efficiency increases with 4.17$\times$ from 2010 to 2018.
This is a power law, so extrapolating this to 12 years gives 6.70$\times$\footnote{I use 12 years instead of 14 from 2009 as typically servers have a
life of 4 years. Therefore the most likely estimate is that the current
servers are two years old, i.e. they have the efficiency from 2021.}. The calculation of the updated emissions per search query is shown
in Alg. \ref{alg:Calculation-of-emissions}.

\begin{algorithm*}
\begin{lyxcode}
PUE:~1.16~in~2010;~1.1~in~2023;~

efficiency~increase~of~hardware~in~12~years:~6.70x~

US~overall~carbon~intensity:~367~gCO\textsubscript{2}/kWh

0.0003{*}(1.1/1.16){*}(1/6.70)~=~0.0000424~kWh~per~search~

0.0000424{*}367~=~0.02~g~CO\textsubscript{2}~per~search
\end{lyxcode}
\caption{Calculation of emissions per search query\label{alg:Calculation-of-emissions}}

\end{algorithm*}

\begin{lyxcode}
\end{lyxcode}
So the energy consumption per conventional search query has dropped
by 7$\times$ in 14 years. There is quite some uncertainty on this
estimate, but it is conservative, so it will not be less than that,
but could be up to 10$\times$. Microsoft has not published similar
figures but there is no reason to assume that their trend would be
different; in fact, their use of FPGAs should in principle lead to
a lower energy consumption per query. In that same period, carbon
emissions per search have dropped about 10$\times$ because of the
decrease in carbon intensity of electricity.

\section{BLOOM energy consumption per query}

In a recent paper \cite{luccioni2022estimating}, Luccioni et al.
analysed the energy consumption per query for the 176B-parameter BLOOM
model, which is of the same complexity as GPT-3.
The model was used to provide AI-generated summaries of search queries.
They measured power consumption over a period of 18 days, in which
the model received an average of 558 requests per hour, for 230,768
requests in total. This resulted 914 kWh of electricity. With the
above figure for electricity carbon intensity, the emissions per query
to the summarising model are shown in Alg. \ref{alg:Calculation-of-emissions-1}. 

\begin{algorithm*}
\begin{lyxcode}
914/230768~=~0.004~kWh/request

US~overall~carbon~intensity:~367~gCO\textsubscript{2}/kWh

367{*}914/230768=1.5~g~CO\textsubscript{2}~per~request~
\end{lyxcode}
\caption{Calculation of emissions per query to the LLM\label{alg:Calculation-of-emissions-1}}

\end{algorithm*}

In other words, the query to the BLOOM model to summarise the search
result costs 75$\times$ more energy than the conventional search
query itself.

\section{ChatGPT energy consumption per query}

There are several estimates of the energy consumption per query for
ChatGPT. I have summarised the ones that I used in the following table.
There are many more, these are the top ranked ones in a conventional
search.

\begin{table}
\begin{centering}
\begin{tabular}{|c|c|c|}
\hline 
Ref & Estimate (kWh/query) & Increase vs conventional search\tabularnewline
\hline 
\hline 
\cite{Sreedhar2023} & 0.001-0.01 & 42-{}-236\tabularnewline
\hline 
\cite{Ludvigsen2023} & 0.0017 - 0.0026 & 40--61\tabularnewline
\hline 
\cite{Zodhya2023} & 0.0068 & 160\tabularnewline
\hline 
\cite{Pointon2023} & 0.0012 & 28\tabularnewline
\hline 
\cite{DeVries2023} & 0.0029 & 68\tabularnewline
\hline 
\end{tabular}
\par\end{centering}
\caption{Estimates of energy consumption per query for ChatGPT}

\end{table}

Reference \cite{DeVries2023} by de Vries uses the estimates from
\cite{Patel2023} for energy consumption but does not present a per-query
value so I used the query estimate from \cite{Patel2023}. Overall,
the estimates lie between 24$\times$ and 236$\times$ (from \cite{Sreedhar2023},
which is a collation of estimates from Reddit and therefore very broad)
or 28$\times$ to 160$\times$ (all other sources). 

I consider any estimate lower than 0.002 kWh/query overly optimistic
and any estimate higher than 0.005 kWh/query overly pessimistic. However,
rather than judging, I calculated the mean over all these estimates.
I used four types of means. Typically, an ordinary average gives more
weight to large numbers; a harmonic mean gives more weight to small
numbers. Given the nature of the data, I think the geometric mean
is the best estimate:

\begin{table}
\begin{centering}
\begin{tabular}{|c|c|}
\hline 
Type of Mean & Mean increase\tabularnewline
\hline 
\hline 
Average & 88\tabularnewline
\hline 
Median & 61\tabularnewline
\hline 
Geometric & 63\tabularnewline
\hline 
Harmonic & 48\tabularnewline
\hline 
\end{tabular}
\par\end{centering}
\caption{Mean increase estimate for ChatGPT queries compared to convential
search queries}

\end{table}

As you can see, there is not that much difference between the geometric
mean and the median. So we can conclude that ChatGPT consumes between
fifty and ninety times more energy per query than a conventional (\textquotedbl Google\textquotedbl )
search, with sixty times being the most likely estimate. These estimates
agree remarkably well with those of the BLOOM model (seventy-five
times). 

\section{Other factors contributing to emissions}

\subsection{Training}

Contrary to popular belief, it is the use of ChatGPT, not its training,
that dominates emissions. I wrote about this in \cite{Vanderbauwhede2023}.
In the initial phase of adoption, with low numbers of users, emissions
from training are not negligible, but in the scenario where conventional
search is replaced by ChatGPT-style queries, which is now the case
for Bing, Google, Yandex, Baidu and many of the less popular search
engines, emissions from training are only a small fraction. How much
is hard to say as we don't know how frequently the model gets retrained
and what the emissions are from retraining as opposed to the original
training; they are almost certainly much lower as the changes in the
corpus are small, so it is tuning.

\subsection{Data centre efficiency}

As far as I can tell, PUE is not taken into account in the above estimates.
For a typical hyperscale data centre, it is around 1.1, so it will
not changes the estimate appreciably.

\subsection{Embodied carbon}

Neither the Google search estimate nor the ChatGPT query estimates
include embodied carbon. The embodied carbon can be anywhere between
20\% and 50\% of the emissions from use, depending on many factors.
My best guess is that the embodied emission are proportionate to the
energy consumption, so this would not affect the factor much.

\section{Conclusion}

Taken all this into account, it is possible that the emissions from
generating AI summaries for search are more than a hundred times that
of the conventional search query. As I don't have enough data to back
this up, I will keep the conservative estimates from above (50$\times$
-- 90$\times$; 60$\times$ most likely for ChatGPT; 75$\times$
for BLOOM).

Now, if we want sustainable ICT, then the sector as a whole needs
to reduce its emissions to a quarter from the current ones by 2040.
The combined increase in energy use and growth in adoption of AI-augmented
search and other generative AI applications is therefore deeply problematic.

\bibliographystyle{IEEEtran}
\bibliography{AI-search-emissions}

\end{document}